\documentclass[pdftex,namedreferences]{SolarPhysics}
\usepackage[optionalrh]{spr-sola-addons} 
\usepackage{graphicx}        
\usepackage{color}           
\usepackage{url}             

\newcommand{\aap}{    {\it Astron. Astrophys.}}

\newcommand{\apj}{    {\it Astrophys. J.}}

\begin{document}

\begin{article}

\begin{opening}

\title{What is the Nature of EUV Waves? First STEREO 3D Observations and Comparison with Theoretical Models}

\author{S.~\surname{Patsourakos}$^{1}$\sep
        A.~\surname{Vourlidas}$^{2}$\sep
        Y.M.~\surname{Wang}$^{2}$\sep
      G.~\surname{Stenborg}$^{3}$\sep 
A.~\surname{Thernisien}$^{4}$}
\runningauthor{S. Patsourakos et al.}
\institute{$^{1}$  Center for Earth Observing and Space Research, College of Science, George Mason University, Fairfax, VA 22030 \\        email: \url{patsourakos@nrl.navy.mil} \\  $^{2}$ Naval Research Laboratory, Space Science Division, Washington,  DC 20375 \\ 
$^{3}$ Interferometrics, Inc, 13454 Sunrise Valley Drive,, Herndon, VA 20171 \\
$^{4}$ USRA, 10211 Wincopin Cir Ste 500, Columbia, MD 2104}

\begin{abstract}

  One of the major discoveries of the Extreme ultraviolet Imaging
  Telescope (EIT) on SOHO were intensity enhancements propagating over
  a large fraction of the solar surface.  The physical origin(s) of
  the so-called `EIT' waves is still strongly debated. They are
  considered to be either wave (primarily fast-mode MHD  waves) or
  non-wave (pseudo-wave)
  interpretations. The difficulty in understanding the nature of EUV
  waves lies with the limitations of the EIT observations which have
  been used almost exclusively for their study. They suffer from low
  cadence, and single temperature and viewpoint coverage. These
  limitations are largely overcome by the SECCHI/EUVI observations
  on-board  the STEREO mission. The EUVI telescopes provide
  high cadence, simultaneous multi-temperature coverage, and two
  well-separated viewpoints. We present here the first detailed
  analysis of an EUV wave observed by the EUVI disk imagers on
  December 07, 2007 when the STEREO spacecraft separation was $\approx
  45^\circ$. Both a small flare and a CME were associated with the
  wave. We also offer the first comprehensive comparison of the
  various wave interpretations against the observations. Our major
  findings are: (1) high-cadence (2.5 min) 171\AA\, images showed a
  strong association between expanding loops and the wave onset and
  significant differences in the wave appearance between the two
  STEREO viewpoints during its early stages; these differences largely
  disappeared later, (2) the wave appears at the active region
  periphery when an abrupt disappearance of the expanding loops occurs
  within an interval of 2.5 minutes, (3) almost simultaneous images at
  different temperatures showed that the wave was most visible in the
  1-2 MK range and almost invisible in chromospheric/transition region
  temperatures, (4) triangulations of the wave indicate it was
  rather low-lying ($\approx$90 Mm above the surface), (5)
  forward-fitting of the corresponding CME as seen by the COR1
  coronagraphs showed that the projection of the best-fit model on the
  solar surface was not consistent with the location and size of the
  co-temporal EUV wave and (6) simulations of a fast-mode wave were
  found in a good agreement with the overall shape and location of the
  observed wave. Our findings give significant support for a fast-mode
  interpretation of EUV waves and indicate that they are probably
  triggered by the rapid expansion of the loops associated with the
  CME.

\end{abstract}
\end{opening}
\section{Introduction}

One of the major discoveries of EIT on SOHO (Delaboudini{\`e}re et
al. 1995) was the existence of conspicuous large-scale EUV propagating
intensity disturbances.  These intensity fronts sometimes have  the
appearance of an almost circular wave front and are frequently called
EIT or EUV waves (e.g., Moses et al. 1997; Thompson et al. 1998;
Thompson et al. 1999).  EUV waves are normally first seen in
close proximity to 
a flaring and erupting active region and they subsequently propagate
over a significant fraction of the visible surface before they become
too faint to be detected.  EIT observations have found that they
travel at speeds 50-400 km/s.  Statistical studies showed that they
are mostly associated with Coronal Mass Ejections (CMEs) and not with
flares (e.g., Biesecker et al. 2002).  There is a tendency for the
fastest EUV waves to be associated with similar phenomena in
H$\alpha$, and soft X-rays and to produce type-II radio bursts (e.g.,
Moreton 1960; Athay and Moreton 1961; Klassen et
al. 2000;  Kahler and Hudson 2001; Cliver at al. 2001;  Narukage et al. 2002; Hudson et al. 2003; Gilbert et
al. 2004; Vr{\v s}nak et al. 2006).  There is an extensive literature on
this subject (e.g.,
Wills-Davey and Thompson 1999;
Delann{\'e}e 2000; 
Klassen et al. 2000; 
Kahler and Hudson 2001;
Wu et al. 2001;
Narukage et al. 2002;
Ofman and Thompson 2002;
Gilbert et al. 2004;
Zhukov and Auch{\`e}re \, 2004;
Podladchikova and Berghmans 2005; 
Warmuth and Mann 2005;
Veronig et al. 2006; 
Vr{\v s}nak et al. 2006).
Reviews on the topic can be
found in Chen and Fan (2005), Gopalswamy et al. (2006), Pick et
al. (2006), Schwenn et al. (2006), Warmuth (2007).

Despite the extensive research on these phenomena, significant
controversy remains over the physical origin(s) of the EUV waves. One
interpretation states that EUV waves are global fast-mode waves
triggered by the associated flare or CME (e.g., Thompson et al. 1999; Wang
2000; Wu et al. 2001; Ofman and Thompson 2002).  
The expectation of finding fast-mode waves in the corona was originally
considered by Uchida (1968).
Fast-mode waves can
travel over large distances over the solar surface since they can
propagate at right angles with respect to the ambient, radial to a
first approximation, magnetic field of the quiet Sun regions which
occupies most of the solar surface, particularly during solar minimum
conditions. Moreover, they have a compressive character which can lead
to the intensity enhancement associated with the EUV waves.  In what
follows we will refer to this interpretation as {\it fast-mode wave}.
A sufficiently impulsive flare and/or CME can generate a wave or
  even a shock-wave which will then give rise to signatures over an
  extended range of wavelengths and physical regimes in the solar atmosphere
  such as, EUV waves, Moreton waves (H$\alpha$), Soft X-rays and type II radio
  bursts.

Another school of thought states that the EUV waves are not waves at
all, but rather the signature of the large scale propagation of the
associated CME. We refer to such interpretations as pseudo-waves.  One suggestion for a
pseudo-wave is the disk projection of the large-scale current shells
enveloping the erupting flux rope of the associated CME \footnote{Flux
  ropes are an integral part of almost all CME models and a large
  fraction of coronagraphic CME observations finds evidence of such
  structures.}  (Delann{\'e}e et al. 2008).  Using a 3-D MHD flux rope model these authors proposed that these
 currents could
be associated with enhanced Ohmic dissipation leading to coronal
heating. This would then  result into the enhanced EUV intensities of the wave
fronts. It is not known whether this enhanced heating is steady
  or impulsive in nature.  In what follows we will refer to this
  interpretation as {\it current-shell} model for brevity.

A hybrid model, with features from both wave and pseudo-wave
  models is the model of Chen et al. (Chen et al. 2002; Chen, Fang and
  Shibata 2005; Chen and Fang 2005 for a review). This model is based
  on the eruption of a flux rope and predicts two types of waves: a
  very fast, super-alfv\'enic piston-driven shock straddling  the
  erupting flux rope which may be identified as the coronal
  counterpart of a Moreton wave and a slower pseudo-wave associated with
successive opening the overlying magnetic field induced by the erupting flux rope.
The latter type of wave is a pseudo-wave, which Chen et al. identify with the EUV waves and is
the result of plasma compression, similar to the conclusion of Delann{\'e}e and Aulanier (1999), who
showed that fast expansion of the magnetic field should compress plasma at the boundaries between expanding
stable flux domains, leading to enhanced emission. We will therefore treat the Chen
  et al. model in tandem with the Delann{\'e}e et al. model since both
  models assert that the EUV waves are formed around the errupting
  flux rope by either enhanced current shells and/or plasma
  compression. Note here that Delann{\'e}e et al. (2008) found that
the disk projection of the current shells produces more contrast than
that from the plasma compression.

Another possibility for a pseudo-wave is successive reconnections
between the expanding large-scale CME with small-scale (cool loops)
quiet Sun structures (e.g, Attrill et al. 2007a,b;
van Driel-Gesztelyi et al. 2008).  These reconnection events will give
rise to enhanced heating which will in turn give rise to the bright
front intensity.  In what follows we will refer to this interpretation as
{\it reconnection fronts}.

Finally, another interesting suggestion regarding the nature of EUV
waves is that they represent {\it MHD solitons} (Wills-Davey, DeForest
$\&$ Stenflo 2007).  For solitary waves the wave speed is dependent on
the pulse amplitude (a function of density enhancement; Wills-Davey,
DeForest $\&$ Stenflo, 2007).  As a general trend, the most
well-defined coronal waves  have higher density enhancement and travel
faster.  More observational signatures from this mechanism need to be
worked out in order to be able to distinguish, for instance,
  between solitary and non-solitary waves.

\begin{figure}    
\centerline{\includegraphics[width=0.9\textwidth,clip=]{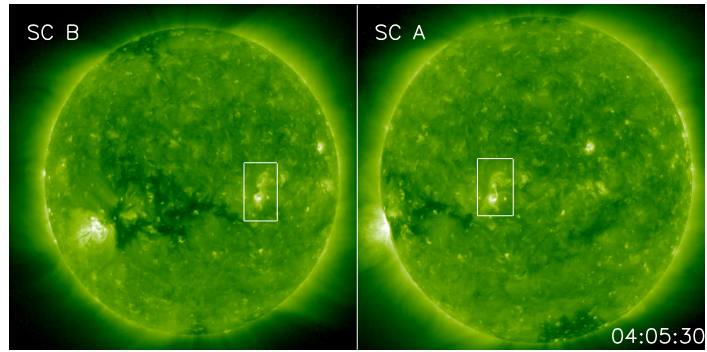}}
\caption{STEREO B (left panel) and A (right panel: full-disk images in
  the 195 channel showing the erupting AR, within the box. The
  separation between the 2 spacecraft was $\approx$ 45
  degrees.}\label{fig:context}
\end{figure}

From the above it is clear that understanding the physical origin of
EUV waves will supply important constraints on coronal conditions
(magnetic field, temperature and density in the corona), for a
fast-mode wave interpretation, or on CME initiation and early
expansion, for a pseudo-wave interpretation. Part of the problem for conclusively
determining the nature of EUV waves stems from the limitations of
the EIT observations used almost exclusively to study these waves:
relatively low cadence (12 min), one temperature (normally the 195
channel) and a single vantage point.

These limitations are largely overcome with the STEREO mission (Kaiser
et al 2008). The Extreme Ultraviolet Imaging Telescope (EUVI; Wuelser
et al. 2004) of the SECCHI instrument suite (Howard et al. 2008)
excels over EIT because of its faster cadence (2.5 min or less
during campaign periods), simultaneous imaging at different
temperature channels, and distinct vantage points allowing a truly 3D
view of the Sun. Indeed, previous analyses of EUV waves with high
cadence EUVI data showed that they propagate at higher speeds than
previously determined, closer to the magnetoacoustic speed, (Long et
al. 2008; Veronig et al. 2008).
Additionally, Gopalswamy et al. (2009)
detected wave reflection from
a coronal hole. Finally, Patsourakos $\&$
  Vourlidas (2007) found evidence of kink-like oscillations of quiet
  Sun loops in the wake of an EUV wave. MHD simulations showed
that EUV waves can induce similar oscillations, at least in the 
source AR loops (Ofman 2007).
All the above findings lend
  strong support to a wave interpretation.

However, these studies relied on a single viewpoint for their
analyses. Here, we present the first stereoscopic study of an EUV wave
which occurred on December 07 2007. Also for the first time, we
  perform a comprehensive comparison of predictions and/or
  expectations from each candidate wave interpretation and test them
  against the STEREO observations. We demonstrate how all the unique
advantages of the STEREO observations place new and strong constraints
on the nature of EUV waves.

\section{Observations and Data Analysis}
We analyzed the observations of an EUV wave seen by the two STEREO
spacecraft (SC; hereafter A and B) from $\approx$ 04:35-05:15 UT
during 7 December 2007.  The STEREO spacecraft were significantly
separated by $\approx 45 ^\circ$. The wave was initiated from the
small active region 10977, which was close to disk center as viewed by
A and towards the East limb as viewed by B (Figure
\ref{fig:context}). The overall configuration of the corona was rather
typical of solar minimum conditions: simple active regions, a
low-latitude coronal hole, with the rest of the disk dominated by
quiet Sun. Close to the time of the wave launch, a small B1.4 class
flare started at around 04:35 in the same AR and peaked at around 04:41
UT. An associated CME was seen later in coronagraph data.  This CME
was rather slow, reaching speeds of $\approx$ 300 km/s in the
coronagraphic fields of view, as determined by {\it CACTUS} (Robbrecht
$\&$ Berghmans 2005) and was not particularly bright.

We focus here on EUV data and total brightness white-light images
collected by the EUVI and the COR1 coronagraph (Thompson et al. 2003)
on SECCHI, respectively.  EUVI takes full disk images in EUV channels
centered around 171, 195, 284 and 304 \AA \, (hereafter referred as
171, 195, 284 and 304).  EUVI has $\approx$ 1.6 arcsec pixels and our
observations have a cadence of 2.5, 10 and 20 minutes for 171, 195-304
and 284, respectively.  COR1 is an internally occulted coronagraph
observing from 1.3-4.0 $R_{\odot}$. The COR1 pixel size is 15 arcsec
and nominal cadence is 15 minutes.  The EUVI and COR1 data were
processed with the standard {\it secchi{\_}prep} routine. Note that
the EUVI and COR1 A \& B image pairs are synchronized; namely, the
observations were taken at the {\it same} time on the Sun. Therefore,
we do not have to worry about the different light travel times from
the Sun to the two STEREO spacecraft.

\begin{figure}    
\centerline{\includegraphics[width=0.9\textwidth,clip=]{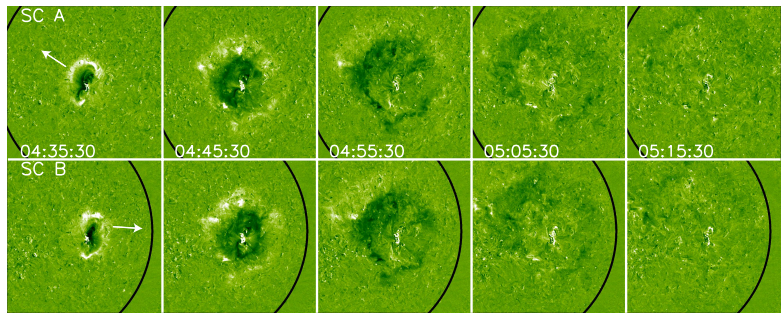}}
\caption{Successive snapshots of the EUV wave in the 195 channel for
  SC A (upper panel) and SC B (lower panel). We display running ratio
  images. All images have the same brightness scale. Ratio increases
  with color from black, to green and white. The arrows in the
    first column panels show the oppositively directed loops observed
    early in the wave event.}
\label{fig:wave_195}
\end{figure}

\subsection{Multi-Viewpoint Temporal Behavior}

A running ratio movie of the EUV wave in the 195 channel for both spacecraft
can be found online (video1.mov) and representative snapshots from
this movie are shown in Figure \ref{fig:wave_195}.  Each frame
corresponds to the ratio of a given 195 image with the previous one.
Note here that the cadence of the 195 EUVI data ($\approx$ 10 minutes)
is similar to that of EIT ($\approx$ 12 minutes): the basic difference
is in the multi-view point aspect of the EUVI observations together
with the higher sensitivity of EUVI compared to EIT.  The latter helps
to better observe faint transients like EUV waves.  Running ratio
  images emphasize changes of the brightness, location, and structure
  of emitting features between two subsequent images and highlight
  faint propagating disturbances such as EIT waves. These images
  correspond to the temporal derivative of the observed intensities.
  The dark regions do not necessarily represent true dimmings
  (e.g. Chertok $\&$ Grechnev 2005).  We provide a
  movie of plain (non-differenced) 195 images (video2.mov), in which
  each frame has been first processed by a wavelet-based technique
  (Stenborg, Vourlidas $\&$ Howard 2007) to increase the image
  contrast.  The propagation of a rather sharp wave-front is evident
  in this movie.

\begin{figure}    
\centerline{\includegraphics[width=0.9\textwidth,clip=]{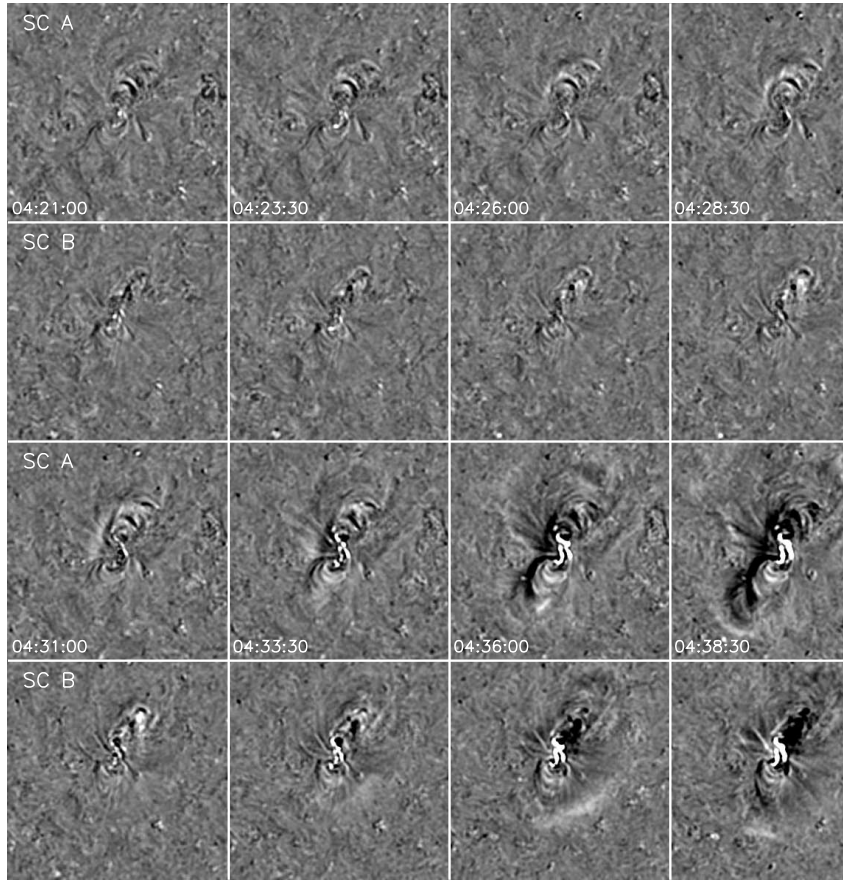}}
\caption{Running ratio wavelet-enhanced images of the target AR taken
  in the 171 channels of A and B with a cadence of 2.5 minutes. Same
  scaling applied to all images.  A 4-frame running ratio is taken,
  i.e. each image is divided by the image taken 10 minutes
  before. Ratio increases with color from black to white.}
   \label{fig:wvl_171}
\end{figure}

The wave first appears at $\approx$ 04:35~UT at the periphery of the
active region, following expanding loop motions which started around
04:15. After 04:35, the wave starts to expand over the solar surface,
becoming progressively more diffuse and faint; after 04:45 it attains
a quasi-circular shape and looks similar to data from both spacecraft.  The wave seems
to avoid the equatorial coronal hole eastwards of the source AR.
By 05:15 the wave covers a significant part of the visible solar
disk, but by this time is rather diffuse and faint. All these characteristics
are pertinent to typical solar minimum EUV waves (e.g., Moses et
al. 1997; Thompson et al. 1998).

\begin{figure}    
\centerline{\includegraphics[width=0.9\textwidth,clip=]{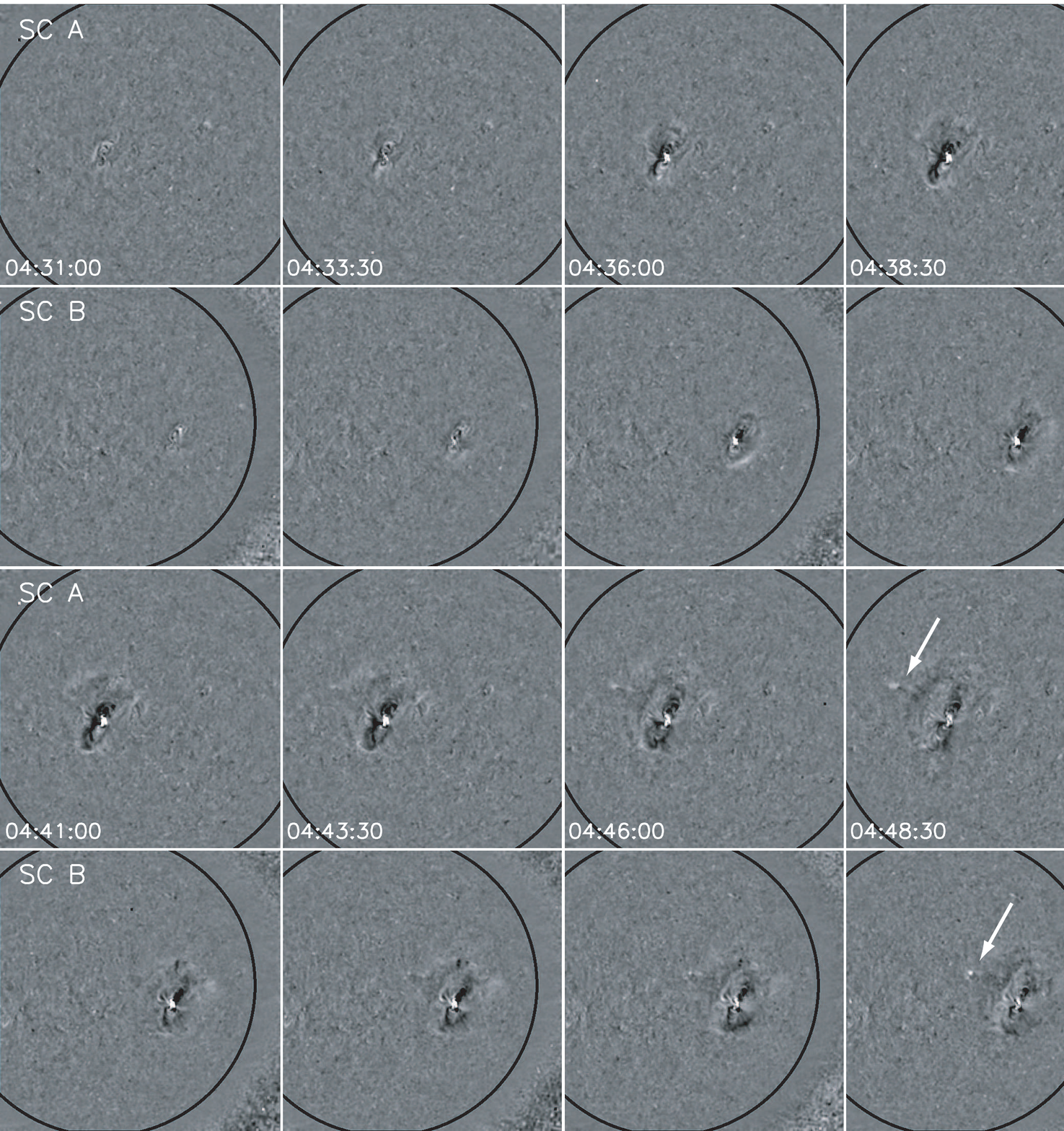}} 
\caption{Running ratio wavelet-enhanced images over an extended area
  taken in the 171 channels of EUVI-A and -B with a cadence of 2.5
  minutes.  A 4-frame running ratio is taken, i.e. each image is
  divided by the image taken 10 minutes before.  The same brightness
  scaling is applied to all images. Ratio increases with color from
  black to white. The arrows indicate a prominence that was
    disrupted by the wave.}
\label{fig:wvl_171_wave}
\end{figure}

However, we noticed a peculiarity in the wave propagation early on:
around 04:35 the wave seems to propagate in opposite directions as
viewed by the two instruments! It propagates eastward in EUVI-A and
westward in EUVI-B (see video1.mov, video2.mov and Figure
\ref{fig:wave_195}).  At this time the wave is incomplete; it is not
forming a full circle.

\begin{figure}    
\centerline{\includegraphics[width=0.9\textwidth,clip=]{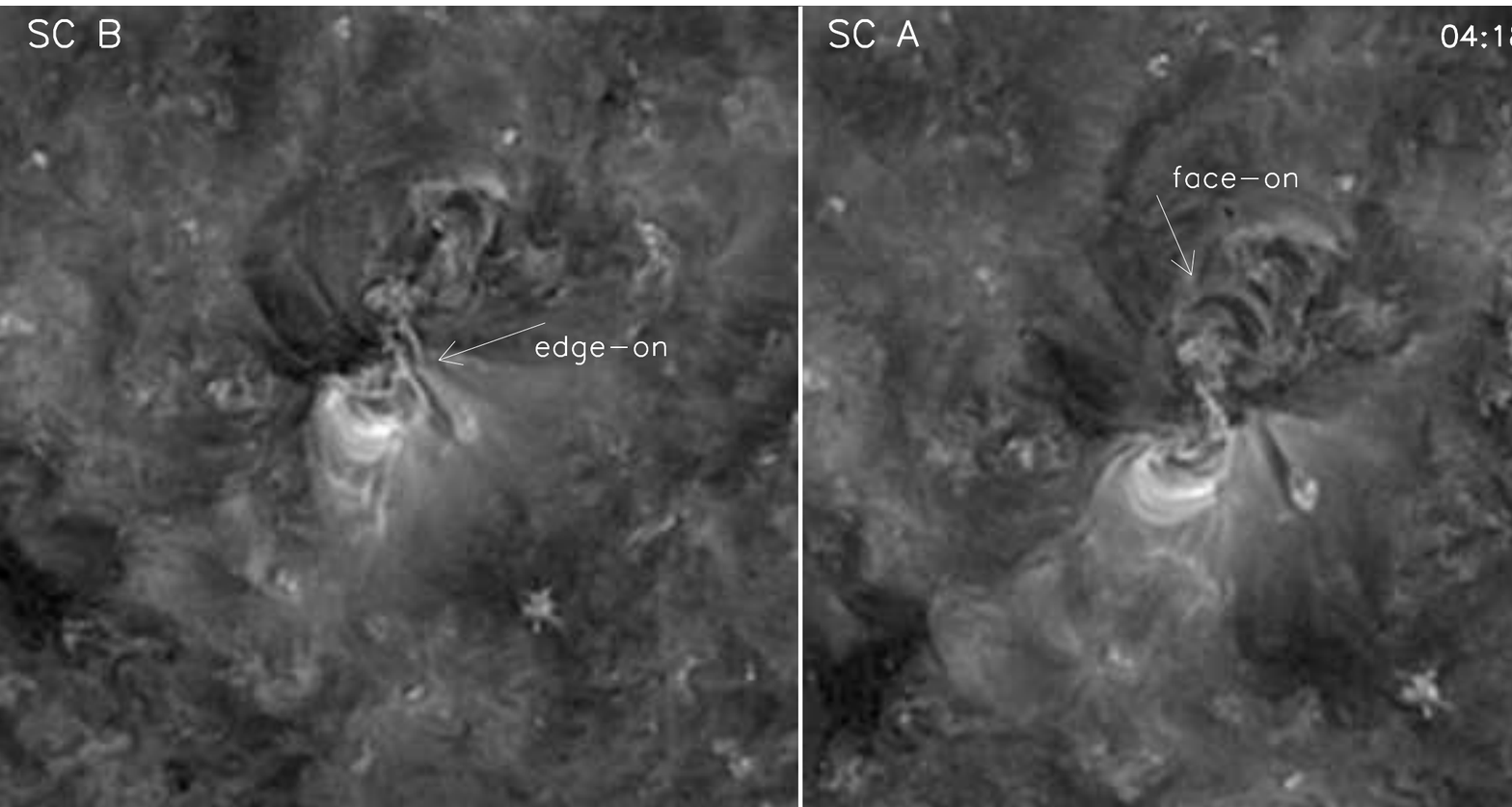}}
\caption{EUVI-A $\&$ B image pair of wavelet-enhanced 171 images
  showing the expanding loops associated with the EUV wave onset. The
  arrows indicate the expanding loops which are seen edge-on and
  face-on in B and A respectively.}
   \label{fig:wvl_171_zoom}
\end{figure}

Higher cadence data are required to resolve this puzzling observation
as well as to establish firmer conclusions about the relative timings
between different aspects of the event (e.g., flare, loop openings,
wave ...).  For this reason, we turn our attention to the 171 images
which were taken at a relatively high cadence of 2.5 min. We process
them with the a wavelet-based technique to increase the image contrast
(Stenborg, Vourlidas $\&$ Howard 2007).  Video3.mov contains an 171 A
\& B movie of the area around the source active region.  Several
snapshots (in running ratio format to emphasize the motions) from the
movie are given in Figure 5.  Video4.mov contains a 171 A \& B movie
over a more extended area, covering a significant fraction of the
solar disk, to capture the large-scale evolution of the wave. Figure
\ref{fig:wvl_171} has several snapshots from this movie in running
ratio format.

Several remarks can now be made. First, loops start to slowly rise
around 04:11 in A and almost 8 minutes later in B (Figure
\ref{fig:wvl_171} and video3.mov).  They are directed towards West
(East) in B (A), similarly to what is seen in the first column of
Figure \ref{fig:wave_195}.  If we now consider the expanding loops in
A indicated by an arrow in the right panel of Figure \ref{fig:wvl_171}
we note that they are seen more or less face-on. Given the relatively large
separation ($\approx 45 ^\circ$) between A and B these loops should be
then seen edge-on in B, as indicated by the arrow in the left panel of
Figure \ref{fig:wvl_171}.  Therefore, the oppositely-directed early
signature of the wave is associated with the expanding motions of the
{\it same} loops observed from the two well-separated
viewpoints. Between $\approx$ 04:31:00-04:33:30 the expanding loops
undergo a sudden jump and cannot be traced anymore; a  wave
front starts to develop (Figure \ref{fig:wvl_171} and video3.mov).
This jump occurs slightly before the start of the associated flare
($\approx$ 04:35).  Given that flare onset and impulsive CME
acceleration seem to simultaneously occur (e.g., Zhang et al.  2001), it is
very possible that the observed loops' jump also marked the the start
of the impulsive acceleration of the associated CME. The first almost
quasi-circular wave front forms around 04:41; the wave also seems to
interact with a prominence to the East of the active region at around
04:43 (Figure \ref{fig:wvl_171_wave} and video2.mov).

The most important findings of this Section, namely that the wave
appeared different, i.e. it exhibited different shapes and
  direction of propagation, during its early phases from the two
distinct vantage points, when loop expansion was clearly observed,
(i.e.  Figures \ref{fig:wave_195}, \ref{fig:wvl_171},
\ref{fig:wvl_171_zoom}) while later on basically the same
quasi-circular pattern was observed have significant
implications. Basically, both pseudo-wave theories expect a
  quasi-circular wave shape {\it at all times} (see Delann{\'e}e et
  al. 2008; Attrill et al. 2007a) which is clearly in disagreement
  with our observations. Furthermore, the fact that during later times
  the wave seems similar on both EUVIs could be a problem for the {\it
    current shell} model of Delann{\'e}e et al. (2008) which expects
  the wave to be the disk projection of a very high altitude current
  shell (280-407 Mm), which should appear significantly different from
  well-separated viewpoints.  This is probably not the case for the
  {\it reconnection fronts} model of Attrill et al. (2007a) which
  assumes that the wave is formed at lower altitudes.

On the other hand, a quasi-circular shape is consistent with a wave
theory, for which the wave is simply the projection of the 3D wave
dome onto the solar surface (e.g. Uchida 1968, Wang 2000). To
  summarize, we find that the {\it transition} of the wave appearance,
  from an asymmetric loop expansion early on to a more symmetric and
  diffuse front, seems to be against the expectations of pseudo-wave
  theories. These theories invoke symmetric loop expansion at all
  times, either via `passive' flux rope eruption or successive
  reconnections. On the other hand, the observed appearance is
  consistent with a wave driven by an impulsive loop expansion.

\begin{figure}    
\centerline{\includegraphics[width=0.9\textwidth,clip=]{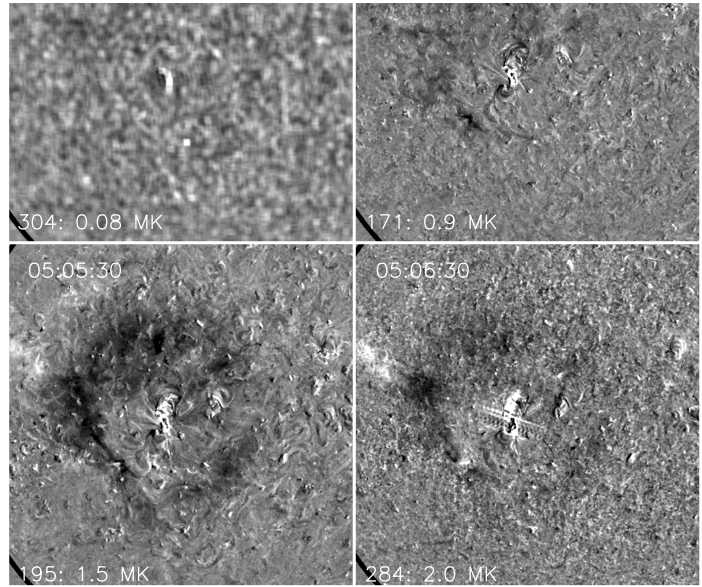}}
\caption{Almost simultaneous snapshots of the EUV wave in the 4
  channels of EUVI-A.  The images were taken between
  05:05:30-05:06:30~UT.  Starting clockwise from the lower left panel
  195, 304, 171, 284. Running ratio images are displayed. All images
  have the same brightness scaling.  Ratio increases with color from
  black to white. Images have been median-filtered by a 3-pixel window
  to reduce noise in all channels except for 304 where we applied a
  15-pixel smoothing to improve the wave visibility. }\label{fig:multi_temp}
\end{figure}

\begin{figure}    
\centerline{\includegraphics[width=0.9\textwidth,clip=]{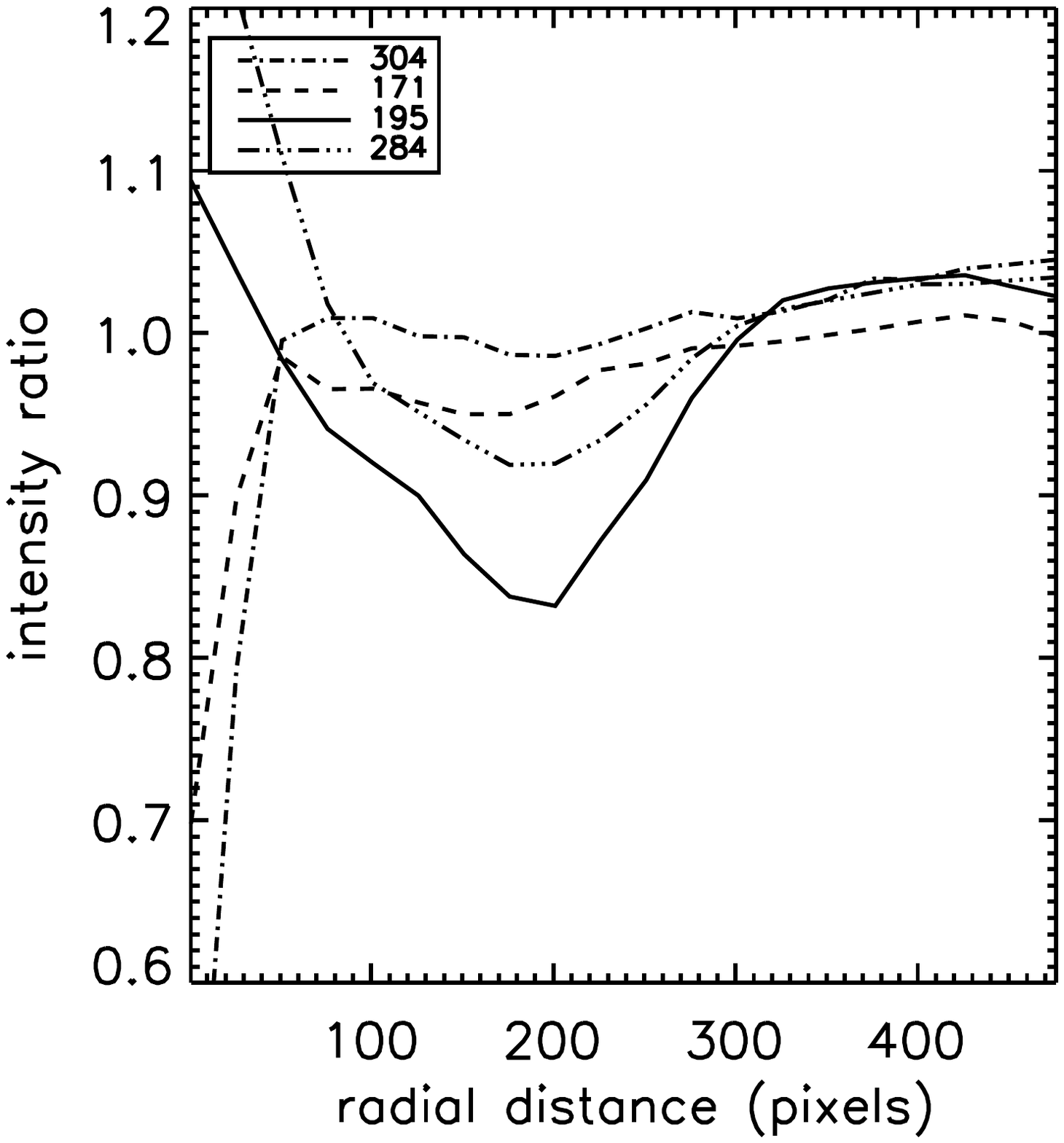}}
\caption{Azimuthally-averaged radial intensity-ratio profiles of the images of \ref{fig:multi_temp}. An EUVI pixel corresponds to $\approx$ 1.6 arcsec.}\label{fig:plot_lena}
\end{figure}

\subsection{Multi-temperature Behavior}
We now study the wave behavior as a function of temperature. Figure
\ref{fig:multi_temp} contains almost simultaneous (within a minute)
running ratio images of the wave seen in all EUVI-A, in
which the wave was closer to disk center compared to EUVI-B. Note that we
used the same 20-minute running ratio images in all channels, which is
the slowest cadence among the EUVI channels. All ratio images have
been scaled to the same ratio range and a spatial 3-pixel wide
median filter has been applied in order to reduce the noise. 
  Azimuthally-averaged radial intensity-ratio profiles of the images
  of Figure \ref{fig:multi_temp} are given in Figure
  \ref{fig:plot_lena}.  They were produced using a procedure similar
  to that described by Podlachikova $\&$ Berghmans (2005): the image
  data were first projected onto a spherical polar coordinates system
  with its center on the flare cite, and we then averaged those maps
  over polar angle to obtain the azimuthally-averaged radial
  intensity-ratio profiles of Figure \ref{fig:plot_lena}. In this plot
  the wave front is located ahead of a region exhibiting a large-scale minimum
  of the intensity ratio (radial distances $>$ 250 pixels).

\begin{figure}    
\centerline{\includegraphics[width=0.9\textwidth,clip=]{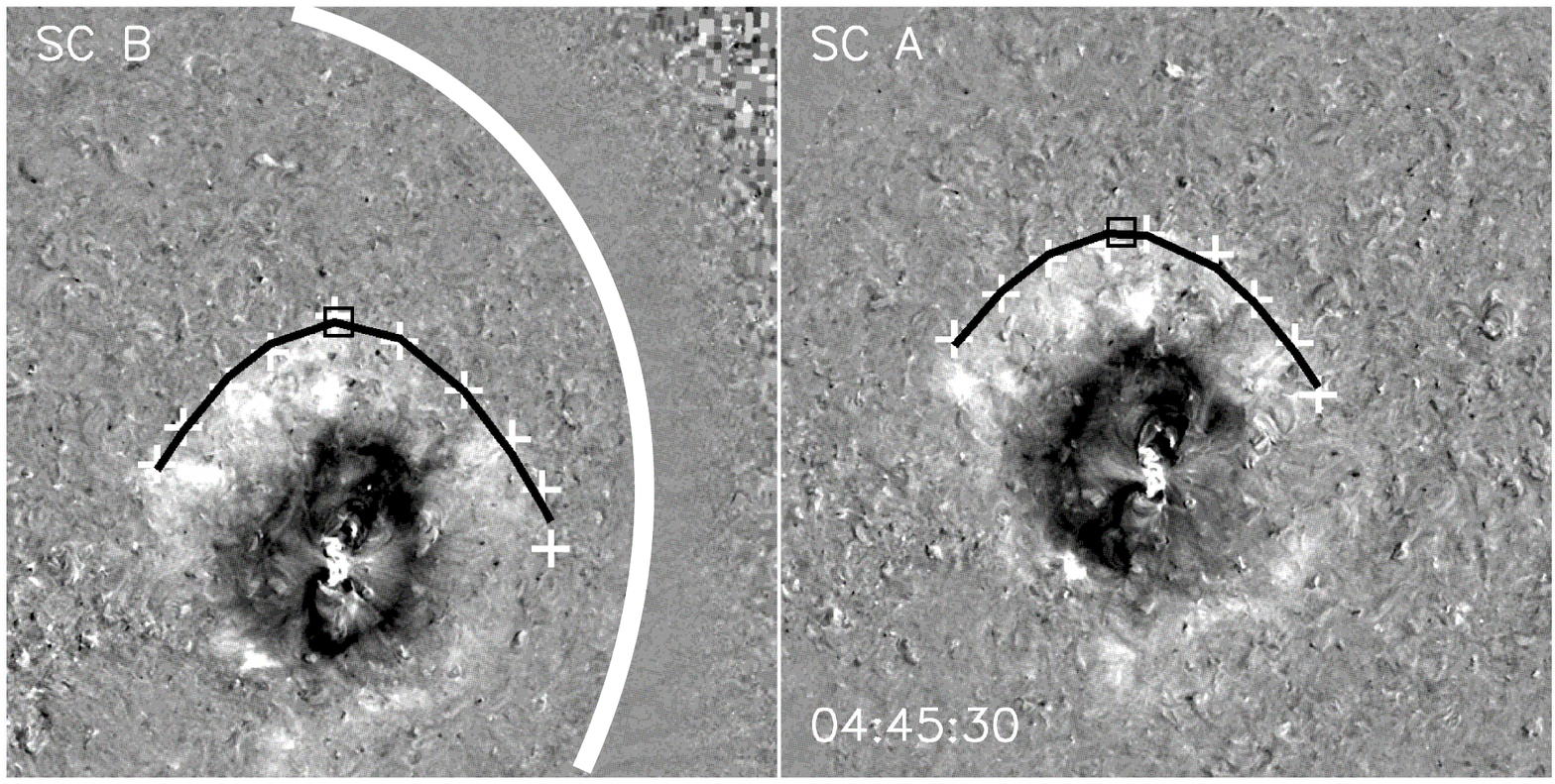}
               }
               \caption{Fitting of segments of the EUV wave front seen
                 in simultaneous EUVI-A and B running ratio images. Both
                 images have the same brightness scaling.  Crosses: points along
                 the wave front selected by point-and-click; solid
                 black line: parabolic fit of the manually selected
                 points. The black boxes are the vertexes of the
                 best-fit parabolas.}\label{fig:triagul}
\end{figure}

As can be seen from Figures \ref{fig:multi_temp} and
\ref{fig:plot_lena} the wave is visible in the coronal channels of
EUVI 171,195 and 284. It is better defined in 195 while its visibility
decreases when going to 284 and 171. On the other hand, the wave is
barely visible in the mainly chromospheric/transition region 304
channel. The wave in this channel was diffuse, and looked
similar to what seen in the 284 channel.  
The intensity increase associated with the wave takes values
in the range of $\approx$ 5-30 $\%$.

The wave propagates over quiet Sun regions, where the bulk 
of the coronal plasma lies in the  temperature range
1-2 MK, with a peak in the quiet Sun Differential Emission
Measure distribution around 1.5 MK (e.g. Brosius et al. 1995), a temperature close
to the peak of the temperature response function of the 195
channel. The fact that the wave is best seen in this temperature range
implies that it represents mostly a density increase
without significant plasma heating taking place. Considering also
the small intensity increase we conclude that
the observed wave represents a small (i.e., linear) density perturbation
of the ambient quiet Sun, without any significant plasma heating taking
place.

This behavior is consistent with coronal propagation of a fast-mode
wave. The wave has a compressive character and when propagating over
the QS corona will compress the ambient plasma, thereby producing
intensity increases which could be the observed wave.  Given the much
higher densities in the chromosphere/transition region it would be
more difficult for a small amplitude wave to produce significant
compression there, which explains the lack of a 304 signature.
  When the eruption is very fast ($>$ several hundred
  $\mathrm{{km}{s}^{-1}}$) a shock wave can be produced.  This will
  give rise to significant density and temperature enhancements over a
  larger temperature range and to sharp wave fronts observed not only in EUV
  but also in chromospheric emissions $H\alpha$ and Soft-Xrays
  (i.e. the `brow' and 'S' wave events) as well type-II radio bursts (e.g. see
  the review of Warmuth 2007).

The observed multi-temperature behavior of the wave poses  problems
to pseudo-wave interpretations.  Consider first the {\it
 current shell} model of Delann{\'e}e et al. (2008).  According to
  this model the wave signature is due to Ohmic heating in large-scale
  current layers at high altitudes occurring around the erupting flux
  rope.  Note here that our best current theoretical understanding of
  coronal heating points to rather small-scale currents as a possible
  heating agent (e.g.  Klimchuk 2006). Moreover, observations show
  little correlation between large-scale currents and SXR intensities
  (Fisher et al. 1998).

  Note now that the density drops-off very rapidly with height in
  the solar atmosphere (it can drop by a factor of $\sim$ 5-10 between the base
  of the corona and 250 Mm; see for example Figure 1.20 in the
  compilation of density measurements of Aschwanden 2005) and the EUV
  intensities are proportional to the line of sight integral of the
  density squared. One would then expect that a disk observation in
  the EUV, such as those of the EUV waves, would be largely dominated
  by the contribution of the lower (and more dense) layers and not
  from the higher layers. However, the {\it current shell} model finds
  that most of the contribution to the current density should
  originate from high-altitudes ($>$ 250 Mm), Delann{\'e}e et
  al. (2008).  The above discussion implies that the {\it current shell} model may not
  lead to any sizeable coronal signal at all.

  We now turn our attention to the {\it reconnection front} model
  of Attrill et al. The observations showed a minimal wave signature
  in the 304 channel.  Note here that the wave did not look
  substantially better in (plain)  304 images processed by
  wavelets.  This operation has  a similar effect 
on the data as differencing (i.e. enhance subtle 
features), but applies to plain images which takes out 
some of the  ambiguities in the interpretation resulting from taking 
differences. Moreover, the 304 data of Figure \ref{fig:multi_temp}
  were largely smoothed by a box-car of 15 pixels to bring up the
  wave. 


  The {\it reconnection front} model of Attrill et al.  relies upon
  small-scale reconnections between the laterally expanding CME and
  small loops in the surrounding QS. The energy content in these
  reconnections would be similar to that associated with the large
  array of transient phenomena known to occur over the QS which are
  widely believed to be the result of small-scale reconnections (see
  the review of Parnell 2002 for example). Therefore, one would expect
  brightenings of similar magnitude in the frame of the Attrill et
  al. model which should be easily observable.  Note here that the
  magnitude of these brightenings would be smaller that those occurring
  in active regions because of the substantially weaker QS magnetic
  fields.  The typical timescales in the
  chromosphere/transition region as well as the lifetimes of these
  transients are in the range 100s-few minutes.  This means that the
  relatively low cadence of the 304 data (10 min), could cause us to
  �miss� the moment of maximum emission of some of the brightenings
  but not all of them. This would give rise to a rather inhomogeneous
  and more importantly incomplete wave-fronts which is probably not
  the case.  Higher-cadence 304 observations are required to show if
  the above is really the case.


The footpoints of coronal structures, where the
bulk of their TR/chromospheric emissions comes from, represent a very
sensitive monitor of coronal conditions.  In the case of steady
  or quasi-steady heating, for example, the TR emission is proportional to the
  coronal pressure (e.g. Klimchuk, Patsourakos and Cargill 2008;
  Equation A12).  Impulsive heating would give rise to sizeable
footpoint emissions, turning them visible in chromospheric/TR
emissions like 304 (Antiochos et al. 2003; De Pontieu et al. 1999;
Hansteen 1993; Klimchuk 2006; Klimchuk et al. 2008; Martens et
al. 2000; Patsourakos and Klimchuk 2008; Spadaro et al. 2006;
Winebarger, Warren and Flaconer 2008).  And this occurs
  irrespectively of the spatial location of the heating along
  individual field lines; thermal conduction and mass flows are very
  efficient at redistributing any excess heat along the field.


  Finally, it is well-known observationally that the quiet Sun
  transition region and chromosphere exhibit stronger variability than
  the overlying corona (e.g.  Berghmans, Clette and Moses 1998;
  Patsourakos and Vial 2002; Teriaca, Madjarska, and Doyle 2002).
  Therefore, if an EUV wave is associated with some sort of impulsive
  heating, one may expect a significant chromospheric/transition
  region signature maybe stronger than the coronal one.  Our
  observations show the contrary.

  The small impact of the wave in the transition region and the
  chromosphere may not be a problem for the {\it current shell}
  model. This is because the heating in that model occurs on much
  higher field lines than for the {\it reconnection front} model. And
  as we saw before footpoint emissions are proportional to the coronal
  pressure, which in turn is also proportional to the field line
  length and therefore height for the same temperature (e.g. the
  scaling-law of Rosner, Tucker and Vaiana 1978).


   Another potential problem for
  pseudo-waves concerns their width.  Assuming a rather low limit for
  the coronal cooling time of 1000 s \footnote{ Coronal oooling times
    depend almost linearly on field line length (e.g., Cargill,
    Mariska and Antiochos 1995) and therefore should increase in the
    expanding field lines associated with a CME and the pseudo-wave.}
  and for a typical EUV wave propagation speed of $\approx$ 300
  $\mathrm{{km}\,{s}^{-1}}$ one finds that a pseudo-wave should result
  in a very extended $\approx$ 3$\times{10}^{5}$ km (or $\approx$ 0.5
  $R_{\odot}$) band of emission in its wake.  This is clearly in
  disagreement with the rather sharply defined wave-fronts observed
  during the early stages of EUV waves (see for example
  video1.mov and particularly video2.mov with the plain wavelet
  images). For instance we estimated a full wave width of   $\approx$
160 arcsec (i.e. $\approx$  110 Mm) for a  snapshot of the wave
taken at 04:45:30; see Figure \ref{fig:wave_195}. This is only the $1/7$ of
$R_{\odot}$.

  We finally comment on the 304 \AA \, channel emission.  Although
  the passband is dominated by the He II line with a peak temperature
  of 0.08 MK (upper chromosphere and lower transition region), this
  passband also contains a contribution from Si XI at 303.32
  \AA \, with a peak formation temperature of $\approx 1.6$ MK which
  is more diffuse than the chromospheric He II emission
  (e.g. Stenborg, Vourlidas, \& Howard 2008). 
  However, spectrally resolved full disk CDS observations
  by Thomson $\&$ Brekke (2000) showed that the Si IX contribution
  in QS areas is very small (4 $\%$ of the He II) and diffuse.
  In our event, the
  wave signatures in the 304 images are diffuse and faint (similar to
  the 284 observations) which suggests a likely coronal origin for
  them.   This is problematic for the {\it reconnection front}
  interpretation where, according to the preceding discussion, we
  should expect low corona/chromospheric signatures \textit{in
    addition} to the coronal signal.

\begin{figure}    
\centerline{\includegraphics[width=0.8\textwidth,clip=]{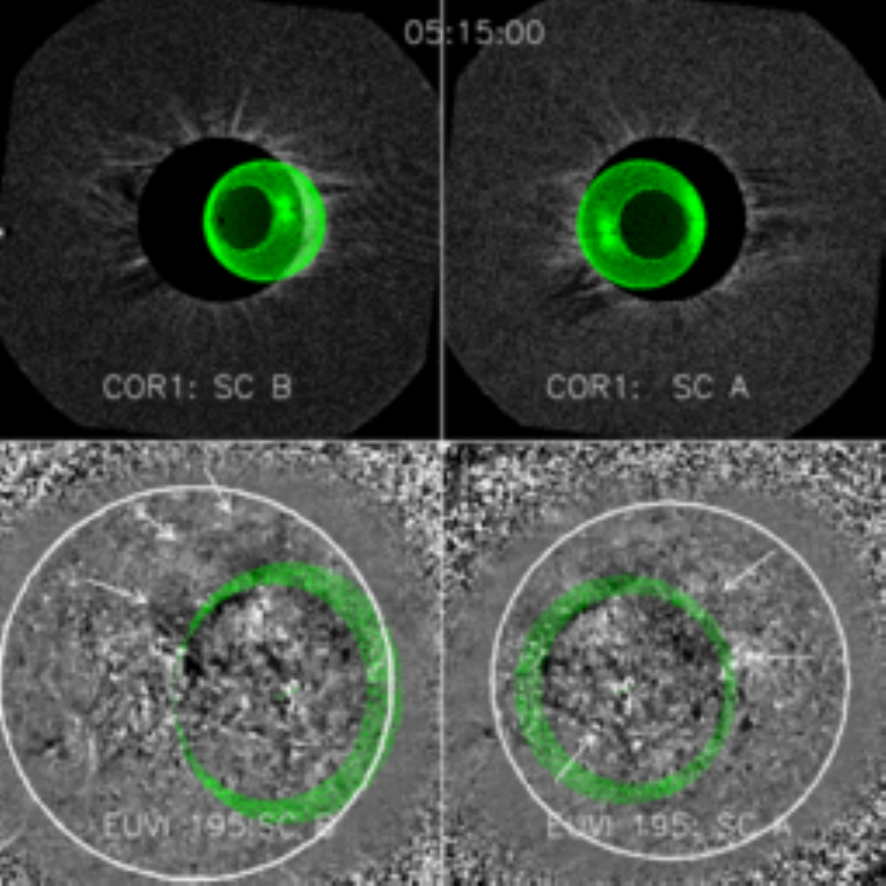}}
\caption{Simultaneous fitting of the CME in COR1-B and A with a flux rope
  model, upper left and right panels respectively. The best-fit
  flux rope model is shown in green.  COR1 total brightness images are
  shown in black and white. Pre-CME images were subtracted.
  Projection (green) of the best-fit flux rope on nearly simultaneous
  EUVI 195 running ratio images for B and A, lower left and right
  panels respectively. We consider rather generous extensions for the wave in the North and
West direction as indicated by the arrows, which takes care of the uncertainties in defining the wave. The South East arrow
indicates the wave front position in that direction. }
\label{fig:fit_cme}
\end{figure}

\subsection{Stereoscopic Analysis}

Thanks to the availability of the two SECCHI viewpoints, we are able to
estimate the 3D location of the wave.  We manually selected several
points along a segment of the wave front in an EUVI running ratio
images pair (Figure \ref{fig:triagul}). The segment covered a
significant fraction of the wave front so that our measurement can be
considered as representative of the structure.  We then fitted these
points with a parabolic function for each telescope and determined the
image x, y coordinates of the corresponding vertexes.  Approximating
the wave front with a parabola is a reasonable but by no means unique
choice: CME fronts almost universally exhibit concave-outwards fronts
and wave fronts can be spherically symmetric.  Triangulation of the
vertex locations using the {\it scc\_triangulate} routine; (see
Inhester (2006) for the basics of triangulation and stereoscopy)
supplied the 3D coordinates (x,y,z) of the vertex of the parabola.

The height $h (= \sqrt{x^{2}+y^{2}+z^{2}}-R_{\odot}$) of the wave
above the solar surface is about 0.13 $R_{\odot}$ or $\approx 90 \pm
7$ Mm (the error is the standard deviation of the heights
determined from 10 repetitions of the above procedure) and is
comparable to the coronal scale-height for quiet Sun temperatures
($\approx$ 70 Mm for a temperature of 1.5 MK).  This is consistent
with fast-mode wave propagation over quiet Sun areas, since the wave
perturbs the ambient coronal plasma with its bulk confined within a
coronal scale-height.

A height of $\approx$ 90 Mm is probably too small for large-scale
current layers around an erupting flux rope of the {\it current shell}
model. Our stereoscopic height observations of the wave of Figure
\ref{fig:triagul} were taken around 30 minutes after loop expansion
was first observed and not very far from when the CME first emerged in
the COR1 field of view (1.5 $R_{\odot} $) which means that the height
of the errupting structure would have been substantially large.
Moreover, Delann{\'e}e et al. (2008) expect that the current
  shells most contributing to the wave should be high-lying ($>$ 250
  Mm).

Finally, our height measurement is probably too large to be consistent
with the {\it reconnection front} model.  The fraction of quiet
  Sun loops which extend to substantially large heights ($>$ 10 Mm)
  represents a small fraction of the QS magnetic flux (10 $\%$
  maximum); with the majority of QS loops not reaching heights larger
  than 10 Mm (Close et al. 2004). In the frame of the Attrill et al
  model,  we expect more reconnections to occur between the
  shorter QS loops and the expanding flux-rope. 
  Also note that
  since the magnetic field strength decays with height, reconnections
  employing low-lying QS loops would be more energetic than those
  employing larger ones. Therefore, the radiative signal of the wave
  in the frame of the {\it reconnection front} model will be dominated
  by the contribution of the low-lying loops. Larger loops could
prevent some of the interactions between the lower loops and the expanding
CME, but possibly not all of them since the small loops are more numerous.

  From the above, and given the lack of any predictions
  on the height of reconnections for the {\it
    reconnection front} model,  we speculate that the wave height in
this scenario would be that of QS cool loops; namely 5-10 Mm tall
(e.g., Aiouaz and Rast 2006; Dowdy, Rabin, and Moore 1986; Feldman,
Widing, and Warren 1999; Close et al. 2004; Patsourakos, Gouttebroze,
and Vourlidas 2007; Peter 2001; S{\'a}nchez Almeida et al.  2007; Tu
et al. 2005).

\begin{figure}    
\centerline{\includegraphics[width=0.8\textwidth,clip=]{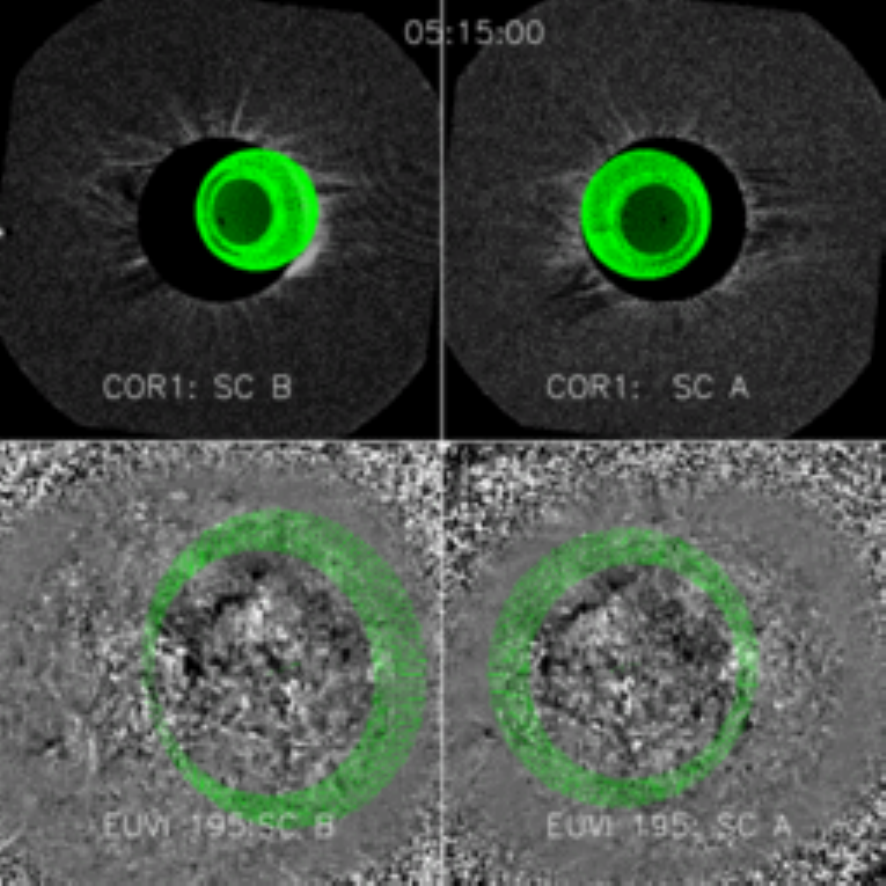}}
  \caption{Determining a flux rope CME model with an on-disk
    projection (green) comparable to the size and location of the
    observed in EUVI 195 wave on both B and A, lower left and right
    panel respectively. Projection (green) of this flux rope model on
    COR1 total brightness images from B and A displayed in black and
    white (upper left and right panel.  respectively). Pre-CME images
    were first subtracted from the COR1 images.  The COR1 and EUVI data
    were taken almost simultaneously. }
   \label{fig:fit_wave}
\end{figure}

\subsection{Pseudo-wave Modeling}
To further test the suggestion that EUV waves could be pseudo-waves directly
linked to the CME expansion, we performed forward modeling of the
associated CME as seen by COR1. 

We used the flux rope CME model (graduated cylindrical shell; GCS) of
Thernisien, Howard and Vourlidas (2006). GCS is a purely
  geometrical 3D model aiming to reproduce the large-scale appearance
  of CMEs in coronagraph data and in particular the CME envelopes: it
  does not aim to reproduce the fine internal structure of CMEs.  It
  makes no assumption about the physical mechanisms producing the
  observed CME envelopes.  The model consists of a tubular section attached
to two cones, which approximate the main body of the flux-rope and its
legs, respectively.   The model is formulated in terms of few
  adjustable parameters, i.e. location, longitude and latitude of the
  source region, height of the CME front, etc (see Thernisien et al
  2006 for details).

  The GCS model makes it easy to calculate projections onto the plane
  of the sky from any viewpoint (e.g. STEREO, SOHO) and compare them
  with coronagraphic CME observations (COR1 in our case); the free
  parameters of the model are modified until the best match is found
  simultaneously in both viewpoints.  This requirement makes the model
  sensitive to changes of its free parameters and therefore results in
  more accurate CME fits than possible with a single viewpoints (Thernisien et al 2009).

We applied the GCS model to an image pair of total brightness COR1
images taken around 05:15 (upper panel of Figure
\ref{fig:fit_cme}). At this time, the wave-associated CME has just
emerged above the COR1B occulter while it was never seen in COR1A
data. Moreover, the EUV wave was still visible on the disk but was
rather diffuse (lower panel of Figure \ref{fig:fit_cme}).  To
construct Fig.~\ref{fig:fit_cme} we did the following:  (1) we
  found the GCS model best reproducing the associated CME observed by
  COR1 (upper panel of Figure \ref{fig:fit_cme}) and, (2) projected
  the upper (i.e. spherical) section of the model onto the co-temporal
  EUVI running ratio images containing the wave (lower panel of Figure
  \ref{fig:fit_cme}). Note that step (1) reproduced the large-scale
  CME seen by COR1-B (top left panel of Figure \ref{fig:fit_cme}), and
  indicated that the CME in COR1-A should lie behind the occulting
  disk (top right panel of Figure \ref{fig:fit_cme}), as it was
  observed.  In step (2) self-similar expansion of the flux-rope like
  CME was assumed as suggested by both observations (Schwenn et
  al. 2003; Cremades $\&$ Bothmer 2004; Thernisien et al. 2006;
  Thernisien et al. 2009) and models (T{\"o}r{\"o}k  and Kliem 2005;
  Attrill et al. 2007a, Delann{\'e}e et al. 2008).

According to pseudo-wave models, the EUV 'waves'
correspond to either the disk projection of the CME envelope for the
{\it current shell model} or to the low coronal lateral extent of the
CME for the {\it reconnection fronts model}.
Therefore, the procedure of Figure \ref{fig:fit_cme}
  allows to validate the realism of pseudo-wave models.  Note here
  that our procedure calculates the projection of the upper section of
  the CME only, and not of its footpoints, which are line-tied to
  their low corona and presumably correspond to areas of strong
  dimmings. In principle, the CME envelope will laterally expand either
  by being `pushed' by the erupting flux rope or by successive
  reconnections with QS loops. In any case, the projection of the CME
  on the low corona---the pseudowave--- will always follow the CME
  evolution. Since the GCS model of Figure \ref{fig:fit_cme} is
  constrained by the position and morphology of the CME in the
  coronagraph data, it accounts for any possible deflections of the
  CME from its original propagation direction (e.g., Thernisien et
  al. 2006).  However, the CME projection disagrees with the actual
  observed wave in two important aspects: (i) it is displaced with
  respect to the wave and, (ii) its projected area is smaller than the
  wave. 

  The displacement between the projected CME and the wave is a strong
  argument towards a wave interpretation (wave and CME propagate
  independently) and against pseudo-wave models (there is a direct
  link between the CME and the wave). The latter would need to invoke
  special conditions to account for such displacements. Besides, the
  often reported CME deviations (due to nearby coronal hole boundaries)  are based on single viewpoint observations (LASCO)
  and therefore may suffer from projection effects. Note here that
  Delann{\'e}e et al. (2008) show a projection of the current shell
  from their model on two snapshots of an observed wave, and a good
  match was found (Figure 11 of Delann{\'e}e et al. (2008)).  However,
  this lacked both the co-temporal snapshot of the associated CME and
  the multiple lines of sights used in our modeling, thereby not
  fully constraining the problem. This underlines the importance of
  multiple viewpoints and of coronagraphs going as close as possible
  to the solar surface.

The size disparity between CME projection and
wave poses another problem for pseudo-wave models
which by definition assume they are equal.
Indeed, if we consider   the very rapid decrease
of density with height and the strong dependence
of EUV intensity on it ($\propto n^{2}dl$, with $dl$ the path
length along the line of sight) 
the signal from
a disk observation of a CME in the EUV  
would be  dominated by the contribution
of its lower sections rather of its 
upper sections. Actually, since CMEs expand with height,
the disk projection of the CME
of Figure \ref{fig:fit_cme} largely overestimates
its actual size which makes the comparison between the observed
wave and pseudo-wave models even more problematic.

  As explained above, the disk projection of the GCS model
  corresponds to the projection of the upper part of CME onto the
  solar disk, and therefore does not correspond to the flux-rope
  footpoints.  In Figure \ref{fig:fit_wave} we do the inverse. We
  first find a flux-rope model with a projection on the disk that
  matches the observed EUV wave simultaneously in A and B.  Then we
  compare its projection on the COR1 images. We believe that this is a
  good method to test the pseudo-wave model prediction; namely that
  the outer CME envelope is the wave. Again, however, we find that the
  pseudo-wave interpretation cannot account for the observations.

 The forward modeling of pseudo-waves of this Section is by no means unique,
in the sense that different geometries and/or eruption scenario
may be employed. However, as explained above, 
the employed geometry (i.e., flux-rope like and self-similar expansion)
find substantial backing in both observations and modeling and therefore
represents a reasonable choice. Finally note here
that if CME and EUV waves had been different facets of the very same phenomenon (i.e. a pseudo-wave), then our fittings 
of Figures 9 and 10 would have been mutually consistent (e.g. the CMEs of Figures 9 and 10 should have been the same), which 
is not the case.

  For our pseudo-wave modeling we had to use co-temporal images of
  the wave and the CME. Because of  its slow speed, the CME took about
  40 minutes after the first appearance of the wave to
  enter the COR1 field of view. By that time the wave had already
  traveled over a significant fraction of the solar area and was
  very diffuse.  We have to wait for observations of more impulsive CMEs to perform CME-wave comparisons, 
with a better defined wave signature.

 \begin{figure}    
\centerline{\includegraphics[width=0.9\textwidth,clip=]{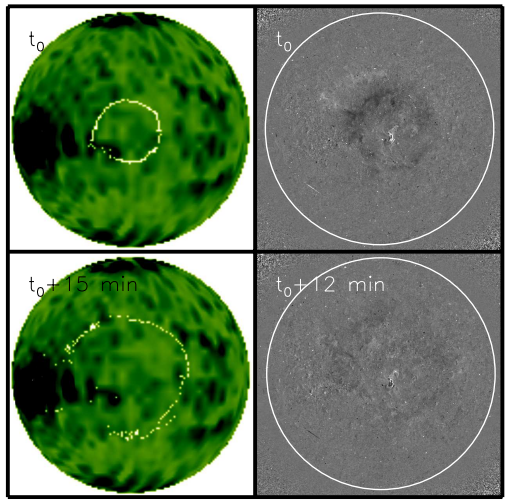}}
\caption{Fast-mode wave simulation of the 2007 December 7 EUV wave (left
  column) and corresponding EIT running ratio images (right column).
  In the left column the location of the wave front projection on the
  solar surface is indicated by white pixels. It is superimposed on
  the distribution of the magnetoacoustic speed at the solar surface:
  black indicates speeds exceeding 500 $\mathrm{km/s}$ while green and
  white indicate lower speeds.  The 2 snapshots from the simulation
  are separated by 15 minutes. A SOHO vantage point is assumed. In the
  right column the 2 EIT snapshots of the wave are separated by 12
  minutes. Both images have the same scaling with ratio increasing
  with color from black to white.}\label{fig:dad}
\end{figure}

\subsection{Fast-mode Modeling}

Having tested the pseudo-wave interpretation against the observations
in the above, we turn to the wave interpretation. We perform a
fast-mode wave simulation for the observed EUV wave following Wang
(2000). First, a Potential Full Source Surface Extrapolation of the
global coronal magnetic field was carried out using MDI synoptic maps
centered at the day of our observations. For each traced field line, a
hydrostatic isothermal atmosphere at 1.5 MK was then attached
following a scaling-law relating density and base magnetic field. The
above steps supplied the 3D distribution of the fast-mode speed in the
corona. Finally, a number of rays have been initiated at the periphery
of the active region, and tracked as a function of time. Essentially
the fast-mode wave is diffracted away from regions with high gradients
in the fast-mode speed.  This happens in active regions (upward wave
diffraction) and in coronal holes (diffraction away from the coronal
hole).  Less diffraction occurs in quiet Sun, where the fast-mode
speed gradients are smoother.  Note that the model makes no explicit
assumption about the physical mechanism triggering the wave.  For more
on the model refer to Wang (2000).

Two snapshots of our simulations are given in Figure 10 where we plot
the projection of the wave front (white dotted line) on the solar
surface. A viewing point from SOHO has been assumed.  The simulated
snapshots are separated by 15 minutes, which is also approximately the
time separation (12 minutes) between the 2 running-ratio EIT images
(Delaboudini{\`e}re et al. 1995) which complement Figure 11.  We see
that the fast-mode simulation essentially manages to track the large
scale appearance of the wave both in terms of size and shape. It
follows rather closely the location and shape of the wave between the
two successive snapshots and even reproduces some details of the wave,
like the avoidance of the coronal hole.

It is interesting to note that if the wave is launched at the center
of the active region, it will expand very rapidly in the radial
direction which is at odds with the observations. The radial expansion
is due to the very rapid decrease of the fast-mode speed with height,
which leads to very strong wave diffraction in the upward
direction. Instead, the observations in section 2.1 show that the wave
is low-lying and appears in the periphery of the active region. 
  This is also a strong argument against the flare as a trigger of the
  wave since the wave would form at the heart of the active region in
  that case.

\section{Summary and Conclusions}

We have performed the first detailed analysis of an EUV wave from two
vantage points as observed by the SECCHI disk imagers and
coronagraphs. The high cadence and stereoscopic observations allowed
us for the first time to precisely determine {\it where} (i.e. in the
AR periphery) and {\it when} (i.e. closely associated with the
impulsive acceleration phase of the CME and the start of the
associated flare) the first signature of the EUV wave is seen. Our
findings place new and tighter constraints on the physics of EUV waves
and are summarized as follows:

\begin{itemize}
\item High-cadence (2.5 min) images showed a strong association
  between expanding loops and the EUV wave onset and significant
  differences in the wave appearance during early stages as viewed by
  the two EUV imagers; these differences largely disappear later.
\item The wave first appears at the AR periphery when an abrupt jump
  of the expanding loops occurs within an interval of 2.5 minutes and
  before the first flare signature.
\item The wave is seen more prominently in emissions formed in the
  $\approx$ 1-2 MK range; little, if anything, is seen in the
  chromosphere/transition region. The wave represents a rather modest
  intensity increase of $\approx 5-30 \%$ over the background.
\item Triangulations showed that the wave was low-lying with a height
  of $\approx$ 70 Mm.
\item The on-disk projection of a geometrical flux rope model of the
  associated CME did not show good agreement with the observed wave; a
  disparity of the size and the location of the projection and of the
  observed wave was found.
\item Simulations of fast-mode wave were found to be in a good agreement
  with the observed wave.
\end{itemize}

The comparison of our findings with the expectations of the different
proposed mechanisms for EUV waves were discussed in the previous
Sections and are summarized in Table-1.  We consider that a fast-mode
wave is the most probable interpretation.  Pseudo-wave theories run
into significant trouble when trying to reconcile several important
elements of the observations: an asymmetric-to-symmetric
  transition in the wave appearance as seen from the two SC, a rather
  poor chromospheric/transition region signature (inconsistent at least
  with the {\it reconnection front} model), probably too small wave
height for the {\it current shell} model and too large for the {\it
  reconnection front} model and a missmatching CME on-disk projection
with respect to the observed wave.  On the other hand, a fast-mode
wave interpretation is consistent with all observations.

  We wish to iterate it here, that by EUV waves we are exclusively
  referring to large scale propagating intensity enhancements, which
  reach significant distances from the source region, especially
  during periods of solar minimum when the coronal landscape is rather
  simple.  We do not exclude the possibility that other propagating
  intensity fronts which follow closely the expanding dimmings could
  be indeed pseudo-waves. However, these dimmings would normally never
  reach the rather global scales of the EUV waves we are considering here. 
Statistical studies of EUV dimming showed an average width
  of only $\approx 36^\circ$ (Reinard and Biesecker 2008).  We think that
  models such as the hybrid model of Chen et al. may be able to
  explain both the dimming and the `true' wave. Further MHD simulations
  mapping the parameter space of different CME-wave-dimming scenarios
  are required.

\begin{center}
\begin{table}
  \caption{Comparison of the EUVI wave observations to the predictions
    of various proposed mechanisms to explain EUV waves. The first
    column gives the physical mechanism, the second column the
    observables and the third column outlines the observations first
    and the corresponding model predictions next.  TR/chromo =
    transition region/chromosphere.}
\begin{tabular}{lll}
\hline
Mechanism & Observable       & Observations/Model Predictions \\
\hline 
              &  {\it multi-viewpoint appearance}         &  initially asymmetric                  \\
               &                                         & and then $\approx$ circular         \\
fast-mode wave    &                                      &   $\approx$ circular after wave is set up                     \\
current shells    &                                       &  significant differences         \\
reconnection fronts &                                     & $\approx$ circular at all times  \\ \hline
              &  {\it multi-temperature}  &        ($\approx$ 1-2 MK)                 \\
fast-mode wave    &                                      &        ($\approx$ 1-2 MK)                 \\
current shells    &                                       &       weak  coronal/TR/chromo        \\
reconnection fronts &                                     & TR/chromo \\ \hline

              &  {\it height}        &     ($\approx$ 90 Mm)                   \\
fast-mode wave    &                                      &   coronal scale-height  ($\approx$ 70 Mm)               \\
current shells    &                                       &   CME height ($> 250$ Mm)       \\
reconnection fronts &                                     & cool loop height ($<$ 10 Mm) \\ \hline

              &  {\it CME projection}  &     offset and smaller (than the wave)         \\
fast-mode wave    &                                      &   N/A              \\
current shells    &                                       &   cospatial and equal     \\
reconnection fronts &                                     &   cospatial and equal \\ \hline

\end{tabular}
\end{table}
\end{center}

We can now draw the following picture about the observed EUV wave.
Coronal loops start to slowly expand (e.g. Figure \ref{fig:wvl_171}
and video3.mov).  This happens clearly before the onset of the flare.
The angular separation of our observations was large enough ( $\approx
45^\circ$) to allow us to observe the {\it same} loops from strikingly
different perspectives early on in the event (edge-on and face on;
first column of Figure \ref{fig:wave_195}, Figure
\ref{fig:wvl_171_zoom}). Just before the start of the
associated flare, the loops abruptly disappear and cannot be traced
anymore; a well-defined  wave front forms at the periphery of the active
region. This front could be generated by a large scale disturbance
induced by either the impulsive CME acceleration or the associated
flare. The latter possibility can be rather safely excluded for
  the following reasons: (1) the associated flare was weak, (2) we
  found in Section 2.5 that launching the wave in the AR core, as
  expected from a flare-associated blast-wave, would not lead to wave
  fronts consistent with the observations.  It is possible that the
  catastrophic energy release that apparently destroys the ascending
  loops within 2.5 minutes is also responsible for the launch of the
  wave which starts to propagate freely. A freely-propagating wave is
  consistent with the fact that the wave becomes more diffuse as it
  propagates farther from the active region (e.g. Parker 1961;
  Hundhausen 1985; Wartmuth 2007).


Before closing we note that clearly more work from both observational
and theoretical approaches is required. First, although the analyzed
event could be considered as a `typical' solar minimum EUV wave, our
study needs to be extended to larger sets of EUV wave events seen by
STEREO. More emphasis should be given to higher cadence wave
observations in the 195 channel, in which waves are best visible, in
order to obtain a larger number of usable measurements per event.
Moreover, higher cadence is needed also in the 304 channel in an
attempt to search for evidence of transient brightenings that could be
associated with reconnection fronts. Combination of stereoscopic wave
observations by STEREO with the ultra-high cadence EUV observations from
SDO will certainly help into that direction.

Finally, we note that the theories of EUV waves have reached various degrees of
sophistication. However none of these theories are yet at a level to
construct synthetic EUV images out of the particular physical
mechanism to allow direct comparison with the observations.  For this
reason, we relied in this study upon morphological concepts like wave
fronts, CME projections, geometrical shapes etc, which have led us to
physical insights but are not sufficiently quantitative, at least to
our satisfaction. We welcome the recent development of full Sun MHD
models of Linker et al. (2008) which treat in a self-consistent manner both
the CME eruption (i.e. the wave driver) and plasma thermodynamics
(i.e. what it ultimately required for calculating realistic EUV
intensities).  The simulated EUV images showed the appearance of
propagating intensity fronts similar to typical EUV waves. The
analysis of these fronts showed they were associated with fast-mode
waves.

\begin{acks}
The SECCHI data used here were produced by an international consortium
of the Naval Research Laboratory (USA), Lockheed Martin Solar and
Astrophysics Lab (USA), NASA Goddard Space Flight Center (USA),
Rutherford Appleton Laboratory (UK), University of Birmingham (UK),
Max$-$Planck$-$Institut for Solar System Research (Germany), Centre
Spatiale de Li\`ege (Belgium), Institut d’ Optique Th\'eorique et
Applique\'e (France), and Institut d’Astrophysique Spatiale (France).
We sincerely thank the referee for many usefull comments that
led to a significant improvement of  the manuscript.
We thank G. Attrill, L. van Driel-Gesztelyi, 
E. Robbrecht, H. Hudson, S. Plunnket, J. Linker, 
S. Krucker and M. Pick
for useful discussions.
\end{acks}


\end{article} 

\end{document}